\documentclass[useAMS,usenatbib]{mn2e}

\usepackage{times}
\usepackage{graphics}
\usepackage{amsmath}
\usepackage{amssymb}
\usepackage{listings}
\usepackage{array}
\usepackage{bm}

\newcommand{\bd}{\begin{displaymath}}
\newcommand{\ed}{\end{displaymath}}
\newcommand{\nbody}[1]{${\scriptstyle NBODY#1}$}
\newcommand{\ZZ}[1]{${\scriptstyle #1}$}
\newcommand{\be}{\begin{equation}}
\newcommand{\ee}{\end{equation}}
\newcommand{\bea}{\begin{eqnarray}}
\newcommand{\eea}{\end{eqnarray}}

\newcommand{\vecR}{{\bf R}}
\newcommand{\vecV}{\dot{\bf R}}
\newcommand{\vecF}{{\bf F}}
\newcommand{\nblist}{{\mathbb L}}

\newcommand{\reg}{{\mathtt R}}
\newcommand{\irr}{{\mathtt I}}
\newcommand{\aarseth}{Aarseth}

\title[Accelerating NBODY6]
{Accelerating NBODY6 with Graphics Processing Units}
\author[Nitadori \& \aarseth]{
	Keigo Nitadori$^1$\thanks{ 
		E-mail:keigo@css.tsukuba.ac.jp (KN);\quad sverre@ast.cam.ac.uk (SJA)
	} and 
	Sverre J.~Aarseth$^2$\footnotemark[1] \\ 
$^1$Center for Computational Science, University of Tsukuba, 1–-1-–1, Tennodai,
Tsukuba, Ibaraki 305-–8577, Japan \\
$^2$ Institute of Astronomy, University of Cambridge, Madingley Road,
Cambridge, CB3 0HA, UK }
\begin{document}
\date{Accepted 2011. Received 2011; in original form 2011 August 1}

\pagerange{\pageref{firstpage}--\pageref{lastpage}} \pubyear{2011}

\maketitle

\label{firstpage}

\begin{abstract}
We describe the use of Graphics Processing Units (GPUs) for speeding up the code
{\ZZ{NBODY6}} which is widely used for direct $N$-body simulations.
Over the years, the $N^2$ nature of the direct force calculation has proved a
barrier for extending the particle number.
Following an early introduction of force polynomials and individual time-steps,
the calculation cost was first reduced by the introduction of a neighbour scheme.
After a decade of GRAPE computers which speeded up the force calculation further,
we are now in the era of GPUs where relatively small hardware systems are highly
cost-effective.
A significant gain in efficiency is achieved by employing the GPU to obtain the
so-called regular force which typically involves some 99 percent of the particles,
while the remaining local forces are evaluated on the host.
However, the latter operation is performed up to 20 times more frequently and may
still account for a significant cost.
This effort is reduced by parallel SSE/AVX procedures where each interaction term is
calculated using mainly single precision.
We also discuss further strategies connected with coordinate and velocity prediction
required by the integration scheme.
This leaves hard binaries and multiple close encounters which are treated by several
regularization methods.
The present \nbody{6\text{--}GPU} code is well balanced for simulations in the particle range
$10^4-2 \times 10^5$ for a dual GPU system attached to a standard PC.
\end{abstract}


\section{INTRODUCTION}

The quest to perform efficient $N$-body calculations has challenged astronomers and
computer scientists ever since the early 1960s.
For a long time progress was slow but so was the increase in computing power.
The first significant advance was achieved by the Ahmad--Cohen (1973, AC) neighbour
scheme which splits the total force into a distant slowly changing part and a local
contribution with shorter time-scale, hereafter denoted as the regular and irregular force.
Considerable progress on the hardware side was made when the GRAPE-type special-purpose
computers were developed in the early 1990s (Makino, Kokubo \& Taiji 1994) and later
improved to GRAPE-4 and GRAPE-6 (Makino et al.~2003).
More recently, the general availability of Graphics Processing Units (GPUs) and the
corresponding CUDA programming language have facilitated large gains in $N$-body
simulations at modest cost.
Early applications based on CUDA (Nyland, Harris \& Prins 2007,
Belleman, Bedorf \& Portegies Zwart 2008)
demonstrated significant speeding-up, approaching GRAPE-6 performance.
Moreover, the introduction of pseudo double precision for the coordinate
differences without sacrificing much efficiency (Nitadori 2009) ensured
increased confidence in the results.
In other problems, the employment of CUDA has already led to Petaflop performance by
combining several thousand GPUs.

In this paper, we are mainly concerned with small stand-alone systems using one or two
GPUs with the code {\ZZ {NBODY6\text{--}GPU}}.
However, mention should also be made of the parallel version {\ZZ {NBODY6^{++}}} which is
capable of reaching somewhat larger particle numbers using several types of hardware
(Spurzem 1999) and is also intended for GPUs.

This paper is organized as follows.
After reviewing some relevant aspects in the standard {\ZZ {NBODY6}} code we discuss
the new treatment of the regular and irregular force.
As a result of improvements in the regular force calculation, the irregular force now
becomes relatively expensive.
Although the latter may also be evaluated on the GPU, the overheads are too large.
Instead we perform this calculation in parallel using SSE
\footnote{For computational terms, see glossary in Appendix A.}
(Streaming SIMD Extensions) and OpenMP in C++ with GCC built-in functions.
The presence of hard binaries requires careful attention, especially because the
regularization scheme involves different precision for the centre of mass (c.m.)
motion as evaluated by FORTRAN and SSE.
A later implementation with AVX accelerated the irregular force calculation.
The performance gain is illustrated by comparison with the basic version at relatively
small particle numbers and we estimate the cost of doing large simulations for
two types of hardware.
Finally, we summarize current experience with small GPU systems and point to
possible future developments.

\section{BASIC NBODY6 CODE}

The code {\ZZ {NBODY6}} was developed during the late 1990s (Aarseth 1999).
It was based on a previous code {\ZZ {NBODY5}} which also employed the AC neighbour
scheme.
Here the main improvement was to replace the fourth-order Adams method by an equivalent
Hermite formulation (Makino 1991) for the case of two force polynomials
(Makino \& Aarseth 1992).
An intermediate step was made with the Hermite individual time-step 
code {\ZZ {NBODY4}} (Aarseth 1996)
designed for use by several generations of GRAPE-type computers.

The combination of Hermite integration with block-steps has proved a powerful
tool in $N$-body simulations.
At earlier times its simple form was beneficial for using together with the
special-purpose GRAPE hardware which supplies the force and its first derivative.
This allows for the construction of an efficient and accurate fourth-order
integration method where the introduction of hierarchical block-steps reduces the
overheads of coordinate and velocity predictions considerably and facilitates
parallel procedures.
Experience has also shown that the Hermite AC block-step scheme is highly cost-effective
in the standard \nbody6 code.



On conventional computers, the regular force calculation dominates the CPU time, with only
a weak dependence on the neighbour strategy.
Thus there are compensating factors when the number of neighbours ($n_i$) is varied.
The new neighbours are chosen at the time of a regular force calculation.
Hence all particles inside the corresponding neighbour radius are selected, together with
any particles in an outer shell approaching with small impact parameter.
Individual neighbour radii $R_{\rm s}$ are adjusted according to the local density contrast,
with additional modifications near the upper and lower boundary.
New values are obtained by the expression
\bea
R_{\rm s}^{\rm new} = R_{\rm s}^{\rm old} \left ( \frac { n_{\rm p}} {n_i} \right )^{1/3} ,
\label{eq:Rs}
\eea
where $n_{\rm p}$ is predicted from the local density contrast.
Note that the case of zero neighbour number which may occur for distant particles is also
catered for.
Formally, explicit derivative corrections to the force polynomial should be carried out
for each neighbour change. However, the same terms are added and subtracted from the respective
polynomials so that this overhead may be omitted, provided the desired results are obtained
at times commensurate with the maximum time-step (Makino \& Aarseth 1992).

The treatment of close encounters forms a large part of the code.
We distinguish between two-body and multiple encounters which are studied by the tools of
Kustaanheimo-Stiefel (1965, KS) and chain regularization (Mikkola \& Aarseth 1993).
Several methods for integrating the KS equations of motion have been used over the years.
The preferred method is a high-order Hermite scheme (Mikkola \& Aarseth 1998) and an
iterative solution without recalculating the external perturbation.
This allows the physical time to be obtained by a sixth-order Taylor series expansion of
the time transformation $t' = R$.
The so-called Stumpff method maintains machine accuracy in the limit of small perturbations
and only requires half the number of steps per orbit compared to the KS fourth-order
polynomial method employed by {\ZZ {NBODY5}}.
Even so, small systematic errors are present over long time intervals.
Hence a rectification of the KS variables $ \bf u, u'$ is performed to be consistent with
the {\it integrated} value of the binding energy (Fukushima 2005).
Further speed-up can be achieved for small perturbations by employing the slow-down
concept in which the time and perturbing force are magnified according to the principle
of adiabatic invariance (Mikkola \& Aarseth 1996).


Given a population of hard binaries, close encounters between binaries and field stars or
other binaries are an interesting feature particularly because collisions or violent ejections
may occur.
The chain concept led to a powerful method for studying strong interactions of 3--5 particles
where two-body singularities are removed (Mikkola \& Aarseth 1993).
Implementation of perturbed chain regularization introduces many complexities, as well as
dealing with internal tidal effects, membership changes or post-Newtonian terms (Aarseth 2003).
As far as the $N$-body code is concerned, the associated c.m. is integrated like a single
particle with the force and first derivative obtained by mass-weighted summation over the
components.
The internal motions are advanced by a high-order integrator (Bulirsch \& Stoer 1966) with
energy conservation better than $10^{-10}$.
Moreover, the solutions for internal KS or chain are continued until the end of the new
block-step before the other particles are treated.
As for KS termination, this is implemented exactly at the end by a simple iteration.
Internal chain integration, on the other hand, is only advanced while the new time
is less than the block-time and any coordinates required by other particles are obtained
by prediction.
Consequently, chain termination is performed at an arbitrary time and a new current
commensurate time is constructed by suitable subdivision.
This in turn reduces the block-step further but is compensated by the relatively small
number of chain treatments.


Finally, in this paper, we omit a discussion of synthetic stellar evolution which is
optional and forms a large part of the code.
Hence for the purpose of GPU developments these aspects may be ignored for simplicity.
In any case, the additional CPU time required by the host is relatively small.

\section{NEW IMPLEMENTATIONS}
We now turn to describing some relevant procedures associated with
the new implementations.
\subsection{Software design}
A subroutine {\tt INTGRT} in \nbody6 drives the numerical integration 
of the AC neighbour scheme.
It uses two libraries during the integration, prepared for the calculation
of the regular and the irregular force, named {\tt GPUNB} and {\tt GPUIRR}.
Although the names of the libraries include {\tt GPU}, non-GPU implementations
exist, e.g. plain C++ versions or high performance versions with
SSE/AVX and OpenMP.
The present \nbody{6\mathchar`-GPU} code employs a GPU version written in CUDA
(with multiple-GPU support) for {\tt GPUNB} and a CPU version with
an acceleration by SSE/AVX and OpenMP for {\tt GPUIRR}.
A first implementation of {\tt GPUIRR} used GPU.
However, it was
slower than a fine-tuned CPU code because of the {\it fine-grained} 
nature of the irregular force calculation.


\begin{figure*}
\begin{center}
\begin{picture}(350, 120)(-10,0)
	\put(  0, 80){\framebox(90, 20){\tt\large intgrt.f}}
	\put(240, 80){\framebox(90, 20){\tt\large GPUNB}}
	\put(240, 10){\framebox(90, 20){\tt\large GPUIRR}}
	\thicklines
	\put( 90, 94){\vector( 1,0){150}}
	\put(240, 86){\vector(-1,0){150}}
	\put( 82, 24){\vector( 1,0){158}}
	\put(240, 16){\line  (-1,0){166}}
	\put( 82, 80){\line  (0,-1){56}} 
	\put( 74, 16){\vector(0, 1){64}} 
	\put(102, 98){$\vecR_j, \vecV_j, m_j, \Delta t_{\reg,j}; \vecR_i, \vecV_i, h^2_i$}
	\put(132,  76){$\vecF_{\reg,i}, \dot\vecF_{\reg,i}, \nblist_i$}
	\put(102,  28){$\vecR_j, \vecV_j, \vecF_j, \dot\vecF_j, t_{\irr, j}, m_j, \nblist_j$}
	\put(132,   6){$\vecF_{\irr,i}, \dot\vecF_{\irr,i}$}
\end{picture}
\end{center}
\caption{A schematic diagram of the \nbody{6\mathchar`-GPU} code.}
\end{figure*}
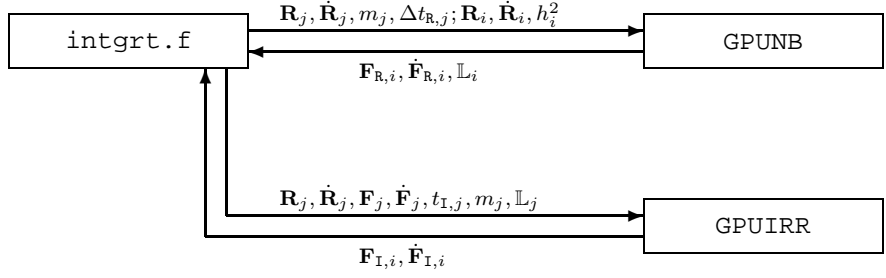

Fig. 1 illustrates a schematic diagram for the relation between {\tt intgrt.f}, {\tt GPUNB}
and {\tt GPUIRR}.
Here,  {\tt GPUNB} receives  positions, velocities, masses and time-steps of the attracting
particles (so called $j$-particles) and positions, velocities and neighbour radii of
the attracted (or active) particles ($i$-particles), and it returns regular forces, their time
derivatives and neighbour lists.
{\tt GPUIRR} holds the position and up to its third time derivatives, mass, 
the time of last irregular integration and the current neighbour list of each particle,
and returns the irregular force and its time derivative for
active particles.
After an irregular step of particle $i$, 
the variables $\vecR_i, \vecV_i, \vecF_i, \dot\vecF_i, t_{\irr, i}$ and  $m_i$
are sent to {\tt GPUIRR} and after a regular step,
the list $\nblist_i$ is sent if there is a change.

\subsection{Velocity neighbour criterion}
The traditional neighbour criterion that a particle $j$ is a 
neighbour of particle $i$ is defined by
\begin{eqnarray}
|\vecR_{ij}| < h_i,
\end{eqnarray}
where $\vecR_{ij} = \vecR_j - \vecR_i$ and $h_i$ is the radius 
of the neighbour sphere.
In this implementation, we employ a modified criterion to 
include the velocities by the condition
\begin{eqnarray}
R_{ij, \rm{min}} \stackrel{\rm def}{=}
\min\left(
	|\vecR_{ij}|, |\vecR_{ij} + \Delta t_{\reg, i} \dot\vecR_{ij}|
\right) < h_i,
\end{eqnarray}
where $\Delta t_{\reg, i}$ is the regular time-step.
Although this increases the computational cost for each pairwise
interaction evaluation in the regular force calculation,
we can take larger regular time-steps for the same number of neighbours.
This safety condition also ensures that high-velocity particles 
can be added to the neighbour list before they come too close.

\subsection{Block-step procedure}
Let us examine the sequential procedure for one block-step.
\begin{enumerate}

\item Obtain the next time for integration,
\begin{eqnarray}
t_{\mathrm{next}} = \min_{1 \le i \le N} (t_{\irr, i} + \Delta t_{\irr, i}),
\end{eqnarray}
where $t_{\irr, i}$ and $\Delta t_{\irr, i}$ are the time of the last
irregular force calculation and irregular time-step of particle $i$.

\item Make the active particle list for regular and irregular force calculation,
\begin{eqnarray}
\nblist_{\mathrm{act}, \reg} = \left\{ i \mid t_{\reg,i} + \Delta t_{\reg,i} = t_{\mathrm{next}} \right\}, \\
\nblist_{\mathrm{act}, \irr} = \left\{ i \mid t_{\irr,i} + \Delta t_{\irr,i} = t_{\mathrm{next}} \right\}, 
\label{eq:Lact}
\end{eqnarray}
where the subscripts $\reg$ and $\irr$ denote regular and irregular terms.
Here, $\left\{ i \mid cond. \right\}$ defines a set of $i$ such that it satisfies $cond.$

\item Predict all particles needed for force evaluation.
\item Calculate the irregular force and its time derivative 
      for particle $i \in \nblist_{\mathrm{act}, \irr}$,
\begin{eqnarray}
{\bf F}_{{\rm I}, i} = \sum_{j  \in \nblist_i}
	m_j \dfrac{\vecR_{ij}}{|\vecR_{ij}|^3}, 
\end{eqnarray}
\begin{eqnarray}
\dot{\bf F}_{{\rm I}, i} = \sum_{j  \in \nblist_i}
	m_j \left[ \dfrac{\vecV_{ij}}{|\vecR_{ij}|^3}  - 3 \dfrac{(\vecR_{ij} \cdot \vecV_{ij})\vecR_{ij}}{|\vecR_{ij}|^5} \right].
\end{eqnarray}

\item Apply the corrector for the active irregular particles
      and decide the next time-step $\Delta t_{\irr, i}$.

\item Accumulate the regular force and its time derivative
      for each active regular particle $i \in \nblist_{\mathrm{act}, \reg}$
      and construct the neighbour list,
\begin{eqnarray}
{\bf F}_{\reg, i} = \sum_{j  \neq i}^{N}
\begin{cases}
	m_j \dfrac{\vecR_{ij}}{|\vecR_{ij}|^3} & (R_{ij,{\rm min}} > h_i) \\
	0                                      & (\text{otherwise})
\end{cases},
\end{eqnarray}

\begin{eqnarray}
\dot {\bf F}_{\reg, i} = \sum_{j  \neq i}^{N}
\begin{cases}
	\displaystyle\scriptstyle{m_j \left[ \frac{\vecV_{ij}}{|\vecR_{ij}|^3}  - 3 \frac{(\vecR_{ij} \cdot \vecV_{ij})\vecR_{ij}}{|\vecR_{ij}|^5} \right]} & 
		\scriptstyle({R_{ij,{\rm min}} > h_i)} \\
	{0}                                   & {(\text{otherwise})}
\end{cases},
\end{eqnarray}
\begin{eqnarray}
\nblist_i = \left\{ j \mid j \neq i, R_{ij, {\rm min}}  < h_i \right\}.
\end{eqnarray}

\item Execute the regular corrector. Since the neighbour list $\nblist_i$
      has been updated, the force polynomials should be corrected to reflect
	  the difference between the old and new list.

\end{enumerate}

\subsection{The {\ttfamily GPUNB} library}

The {\tt GPUNB} library computes regular forces and creates neighbour lists
for a given set of particles with single or multiple GPU(s).
First, the predicted position, velocity and mass of $N_j$ particles are sent
from the host. The regular force and neighbour lists
of $N_i$ particles are then computed. Therefore, $N_i N_j$ pairwise interactions
are evaluated in one call.
Unless $N_i$ is much smaller than $N_j$, the cost of the prediction and data transfer
is not significant.

The basic method to calculate the force on GPUs is common to the previously
existing implementations (Nitadori 2009, Gaburov, Harfst \& Portegies Zwart 2009),
except for the treatment of neighbours.
During the  accumulation of forces from all $N_j$ particles, when $j$ is a neighbour,
{\tt GPUNB} skips the accumulation and saves the index $j$ in the neighbour list.
This procedure is applied for all $N_i$ particles in parallel, using the 
many {\it threads} of the GPU.

Now we discuss in more detail the force calculation procedure of {\tt GPUNB}.
Each $i$-particle is assigned to each thread of GPU.
The position, velocity, neighbour radius, force, its time derivative and 
neighbour count of particle $i$ are held on the registers of each thread.
The position, velocity and mass of particle $j$ are broadcast from the
memory to all the threads in a thread-block, and forces on multiple $i$-particles
are evaluated in parallel.
If particle $j$ turns out to be a neighbour of $i$, the index $j$ is
written to the neighbour list in the memory.

The actual behaviour of each thread is given in Listing B2 with the 
CUDA C++ language. There is an `if statement' for the neighbour treatment,
which is translated into a mask operation by the compiler and has little
impact on performance. If the branch is not removed, it would be a 
serious overhead for parallel performance.

As well as the $i$-parallelism for multiple threads, $j$-parallelism
for multiple thread-blocks and multiple GPUs are also exploited.
Thus, after the force computations, partial forces and partial neighbour
lists of certain particles are distributed for multiple thread-blocks.
To minimize the data transfer from GPU to host, we prepared kernels
for force reduction and list gathering, where local indices in 16-bit integer
are translated to global indices in 32-bit integer.
Partial forces are summed and sparsely scattered partial lists are
serialized to a linear array before they are sent back to the host PC.

Finally, single precision arithmetic turned out to provide sufficiently
accurate results for practical use. 
This is because all the close interactions are skipped in the regular
force procedure 
\footnote{We still left options using the `two-float' method for the coordinates
and the accumulator.}.

\subsection{The {\ttfamily GPUIRR} library}

The name of the library {\tt GPUIRR} comes from an early effort 
to accelerate the irregular force calculation on GPU, although the
GPU is not used in the current implementation,
just because it is slower than a well tuned
CPU code.
Still, the name of the library and its API are used in the fast
version on CPU with SSE/AVX and OpenMP.

Different from {\tt GPUNB}, {\tt GPUIRR} retains many internal states.
The position and up to its third derivative, mass and time of the last
integration of particle $i$ are held to construct the predictor.
These quantities are updated after each irregular step.
Additionally, the neighbour list of particle $i$ is saved for the computation
of the irregular force. The list is copied from the host routine {\tt intgrt.f}
after each regular step.
When an irregular force on particle $i$ is requested,
the library calculates $\vecF_{\irr, i}$ and $\dot\vecF_{\irr, i}$.
Actually, it can be performed in parallel, i.e. the library receives
a list of irregular active particles $\nblist_{\mathrm{act}, \irr}$ and returns an array
of the force and its time derivative.

The library is tuned for multi-core CPUs and SIMD instructions such as SSE and AVX.
An OpenMP parallelization for different $i$-particles is straightforward.
On the other hand, the SIMD instructions are exploited for the $j$-parallelism
for one $i$-particle, which requires technical coding.
To calculate pairwise forces on a particle $i$ from multiple 
(4-way for SSE, 8-way for AVX) $j$-particles, we need to gather them
from non-contiguous addresses indicated by the neighbour list.
The gathering process is relatively expensive and the GFLOPS rating or
`number of interactions per second' is decreased  to about 40 percent of
the case of brute force calculations (Tanikawa et al. 2012)
on the same processor with the same AVX instruction set.

Unlike the regular force calculations, full single precision calculation
of the irregular force may not be accurate enough during  close encounters. 
Thus, we employ the so-called `two-float' technique to express the
coordinates of each particle (Nitadori 2009, Gaburov et al. 2009).
For the summation of force, single precision turned out to be sufficient
since we do not accumulate too many terms (more than a few hundred) in the irregular
force calculation.

Prediction of positions and velocities are performed inside the library.
When $N_{\mathrm{act},\irr}$ is small, predicting all the $N$ particles is not efficient.
Thus, at some point (cf. {\tt NPACT} in Table C1), 
we switch to predict only the necessary particles in this block-step. 
Even if we only predict the necessary particles, combining and sorting of
the neighbour lists or random memory access may be expensive.
Thus there is a turn-around point, and an example is shown in 
the lines 24-29 in Listing C1.

\subsection{Binary effects}
The presence of binaries poses additional complications when combining results of
force calculations from the host and the library for irregular force.
As can be seen above, the irregular force on a single particle due to other
single particles is of simple form and its calculation can be speeded up in C++
with SSE/AVX and OpenMP.
This is no longer the case for regularized systems where differential force
corrections are needed on the host to compensate for the c.m. approximation
which is employed.
For simplicity we assume that only the irregular force needs to be corrected;
this in turn implies that the neighbour radius is sufficiently large.
A further complication arises in the force evaluation due to perturbers.
The numerical problem can be illustrated by considering the new regular
force difference during the corresponding time interval $t - t_{\reg,i}$,
\begin{eqnarray}
\Delta{\bf F}_{\reg} = 
    ({\bf F}^{\rm new}_{\reg} - {\bf F}^{\rm old}_{\reg}) 
  + ({\bf F}^{\rm new}_{\irr} - {\bf F}^{\rm old}_{\irr}),
\end{eqnarray}
where 
${\bf F}^{\rm old}_{\reg}$ and ${\bf F}^{\rm new}_{\reg}$
denote regular forces evaluated at the old and new times
$t_{\reg,i}$ and $t$, respectively, while
${\bf F}^{\rm old}_{\irr}$ and ${\bf F}^{\rm new}_{\irr}$
denote irregular forces both evaluated at the new time $t$
with the old and new neighbour lists.
Hence the net change of irregular force is contained within the second bracket.
In the case of no neighbour change, this term should be suitably small, otherwise
it would tend to reduce the regular time-step.
It is therefore important to ensure consistency between the irregular force as
calculated at each time-step and that obtained elsewhere at the end of a regular step
where both are based on the same {\it predicted} quantities.

In the case of an active KS binary, the old and new force contributions due to perturbers
are acquired in double precision on the host and similarly for the c.m. term which is
subtracted.
Since the latter is evaluated in the single precision {\tt GPUIRR} library, this means that
a small discontinuity is introduced.
However, the old and new irregular force are still numerically identical in the absence
of neighbour change and hence the same neighbours do not affect the regular force
difference.
A similar differential force procedure is carried out for single particles having perturbed
binaries as neighbours.
Note that the sequential ordering of neighbour lists facilitates the identification of KS
binaries for various purposes.
Finally, force corrections involving a chain c.m. particle are performed analogously,
where the internal contributions to $\vecF$ and $\dot\vecF$ are obtained by summation
over the members.

The strategy for controlling the number of neighbours also changes when using the GPU.
Unless present initially, binaries eventually become an important feature of $N$-body
simulations.
It is therefore desirable to strike a balance between the maximum size of regularized
binaries and the neighbour radius.
This can readily be done for long-lived binaries, subject to the maximum membership denoted
by the parameter {\tt {NBMAX}}.
The average neighbour number can also be controlled to a certain extent.
This is achieved by an optional adjustment of a density contrast parameter $\alpha$ used
in the predicted neighbour number $n_{\rm p}$ of equation (\ref{eq:Rs}).
Fortunately the ratio $R_{\rm s} / a$ becomes more favourable for large $N$ because
the semi-major axis of hard binaries is $\propto 1/N$ while the neighbour radius reduces
more slowly.
Hence a safe strategy for small and intermediate simulations is to employ a relatively large
value of {\tt {NBMAX}} together with fine-tuning of $\alpha$.
Also note that the possible problem of too small neighbour radii is mostly relevant for the central
cluster region because $R_{\rm s}$ tends to be larger at lower density.

\section{PERFORMANCE}

The acid test of any code is measured by its performance.
This is especially the case for $N$-body simulations where an important challenge
is to describe the dynamics of globular clusters.
Here we report on some performance tests to illustrate the calculation cost for a
wide range of particle numbers.
It should be emphasized that timings based on idealized systems do not demonstrate
the capability of dealing with more advanced stages which are characterized by
large density contrasts and the presence of hard binaries.
Additional examinations of more realistic conditions are therefore highly desirable,
as commented on below.

\begin{table}
\caption{Hardware and software configurations.}
\begin{tabular}{lll}
\hline
 & System A & System B \\
\hline
CPU         & Core i7--920          & Core i7--2600K           \\
\, (spec)   & \, 4 cores, 2.66 GHz & \, 4 cores, 3.40 GHz    \\
\, (SIMD)   & \, SSE4.2 (4-float)  & \, AVX (8-float)    \\
GPU         & $2 \times$GeForce GTX 470 & $2\times$GeForce GTX 560 Ti \\
Motherboard & MSI X58              & MSI P67                 \\
Memory      & 6 GB, DDR3--1333      & 8 GB, DDR3--1600         \\
OS          & CentOS 5.5 x86\_64   & CentOS 6.0 x86\_64      \\
Compilers   & GCC 4.1.2, CUDA 3.0  & GCC 4.6.1, CUDA 4.0     \\
\hline
\end{tabular}
\end{table}

\begin{table}
\begin{center}
\caption{
	Performance summary for System B.
	We show total wall-clock time and
	partial time for regular and irregular force.
	The number of regular and irregular individual steps 
	is also given, together with the time-step weighted average
	neighbour numbers.
}
\begin{tabular}{lllll}
\hline
$N$                              & 32k  & 64k  & 128k  & 256k \\
\hline
$T_{\rm wall} / {\rm sec}$       & 18   & 59   & 209   & 802   \\
$T_{\rm reg\ }  / {\rm sec}$     & 5.3  & 19   &  84   & 338   \\
$T_{\rm irr\ }  / {\rm sec}$     & 4.7  & 13   &  47   & 230   \\
${\mathcal N}_{\rm reg} / 10^6$  & 2.8  & 6.1  & 13    & 28    \\
${\mathcal N}_{\rm irr} / 10^6$  & 33   & 85   & 220   & 550   \\
${\mathcal N}_{\rm irr} / {\mathcal N}_{\rm reg}$  & 11.8  & 13.9  & 16.9    & 19.6   \\
$\langle N_{\rm nb} \rangle$     & 46   & 55   & 64    & 83    \\
\hline
\end{tabular}
\end{center}
\end{table}

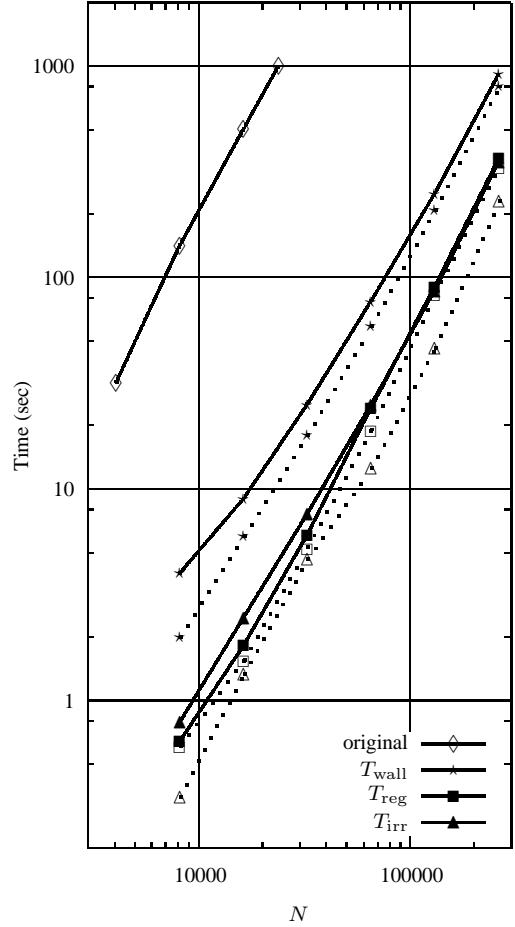
\begin{figure}
\setlength{\unitlength}{0.240900pt}
\ifx\plotpoint\undefined\newsavebox{\plotpoint}\fi
\sbox{\plotpoint}{\rule[-0.200pt]{0.400pt}{0.400pt}}%
\begin{picture}(826,1445)(0,90)
\sbox{\plotpoint}{\rule[-0.200pt]{0.400pt}{0.400pt}}%
\put(150.0,205.0){\rule[-0.200pt]{2.409pt}{0.400pt}}
\put(805.0,205.0){\rule[-0.200pt]{2.409pt}{0.400pt}}
\put(150.0,337.0){\rule[-0.200pt]{2.409pt}{0.400pt}}
\put(805.0,337.0){\rule[-0.200pt]{2.409pt}{0.400pt}}
\put(150.0,405.0){\rule[-0.200pt]{2.409pt}{0.400pt}}
\put(805.0,405.0){\rule[-0.200pt]{2.409pt}{0.400pt}}
\put(150.0,437.0){\rule[-0.200pt]{160.198pt}{0.400pt}}
\put(150.0,437.0){\rule[-0.200pt]{4.818pt}{0.400pt}}
\put(130,437){\makebox(0,0)[r]{ 1}}
\put(795.0,437.0){\rule[-0.200pt]{4.818pt}{0.400pt}}
\put(150.0,537.0){\rule[-0.200pt]{2.409pt}{0.400pt}}
\put(805.0,537.0){\rule[-0.200pt]{2.409pt}{0.400pt}}
\put(150.0,669.0){\rule[-0.200pt]{2.409pt}{0.400pt}}
\put(805.0,669.0){\rule[-0.200pt]{2.409pt}{0.400pt}}
\put(150.0,737.0){\rule[-0.200pt]{2.409pt}{0.400pt}}
\put(805.0,737.0){\rule[-0.200pt]{2.409pt}{0.400pt}}
\put(150.0,769.0){\rule[-0.200pt]{160.198pt}{0.400pt}}
\put(150.0,769.0){\rule[-0.200pt]{4.818pt}{0.400pt}}
\put(130,769){\makebox(0,0)[r]{ 10}}
\put(795.0,769.0){\rule[-0.200pt]{4.818pt}{0.400pt}}
\put(150.0,870.0){\rule[-0.200pt]{2.409pt}{0.400pt}}
\put(805.0,870.0){\rule[-0.200pt]{2.409pt}{0.400pt}}
\put(150.0,1002.0){\rule[-0.200pt]{2.409pt}{0.400pt}}
\put(805.0,1002.0){\rule[-0.200pt]{2.409pt}{0.400pt}}
\put(150.0,1070.0){\rule[-0.200pt]{2.409pt}{0.400pt}}
\put(805.0,1070.0){\rule[-0.200pt]{2.409pt}{0.400pt}}
\put(150.0,1102.0){\rule[-0.200pt]{160.198pt}{0.400pt}}
\put(150.0,1102.0){\rule[-0.200pt]{4.818pt}{0.400pt}}
\put(130,1102){\makebox(0,0)[r]{ 100}}
\put(795.0,1102.0){\rule[-0.200pt]{4.818pt}{0.400pt}}
\put(150.0,1202.0){\rule[-0.200pt]{2.409pt}{0.400pt}}
\put(805.0,1202.0){\rule[-0.200pt]{2.409pt}{0.400pt}}
\put(150.0,1334.0){\rule[-0.200pt]{2.409pt}{0.400pt}}
\put(805.0,1334.0){\rule[-0.200pt]{2.409pt}{0.400pt}}
\put(150.0,1402.0){\rule[-0.200pt]{2.409pt}{0.400pt}}
\put(805.0,1402.0){\rule[-0.200pt]{2.409pt}{0.400pt}}
\put(150.0,1434.0){\rule[-0.200pt]{160.198pt}{0.400pt}}
\put(150.0,1434.0){\rule[-0.200pt]{4.818pt}{0.400pt}}
\put(130,1434){\makebox(0,0)[r]{ 1000}}
\put(795.0,1434.0){\rule[-0.200pt]{4.818pt}{0.400pt}}
\put(150.0,1534.0){\rule[-0.200pt]{2.409pt}{0.400pt}}
\put(805.0,1534.0){\rule[-0.200pt]{2.409pt}{0.400pt}}
\put(150.0,205.0){\rule[-0.200pt]{0.400pt}{2.409pt}}
\put(150.0,1524.0){\rule[-0.200pt]{0.400pt}{2.409pt}}
\put(192.0,205.0){\rule[-0.200pt]{0.400pt}{2.409pt}}
\put(192.0,1524.0){\rule[-0.200pt]{0.400pt}{2.409pt}}
\put(224.0,205.0){\rule[-0.200pt]{0.400pt}{2.409pt}}
\put(224.0,1524.0){\rule[-0.200pt]{0.400pt}{2.409pt}}
\put(250.0,205.0){\rule[-0.200pt]{0.400pt}{2.409pt}}
\put(250.0,1524.0){\rule[-0.200pt]{0.400pt}{2.409pt}}
\put(272.0,205.0){\rule[-0.200pt]{0.400pt}{2.409pt}}
\put(272.0,1524.0){\rule[-0.200pt]{0.400pt}{2.409pt}}
\put(292.0,205.0){\rule[-0.200pt]{0.400pt}{2.409pt}}
\put(292.0,1524.0){\rule[-0.200pt]{0.400pt}{2.409pt}}
\put(309.0,205.0){\rule[-0.200pt]{0.400pt}{2.409pt}}
\put(309.0,1524.0){\rule[-0.200pt]{0.400pt}{2.409pt}}
\put(324.0,205.0){\rule[-0.200pt]{0.400pt}{320.156pt}}
\put(324.0,205.0){\rule[-0.200pt]{0.400pt}{4.818pt}}
\put(324,164){\makebox(0,0){ 10000}}
\put(324.0,1514.0){\rule[-0.200pt]{0.400pt}{4.818pt}}
\put(424.0,205.0){\rule[-0.200pt]{0.400pt}{2.409pt}}
\put(424.0,1524.0){\rule[-0.200pt]{0.400pt}{2.409pt}}
\put(483.0,205.0){\rule[-0.200pt]{0.400pt}{2.409pt}}
\put(483.0,1524.0){\rule[-0.200pt]{0.400pt}{2.409pt}}
\put(524.0,205.0){\rule[-0.200pt]{0.400pt}{2.409pt}}
\put(524.0,1524.0){\rule[-0.200pt]{0.400pt}{2.409pt}}
\put(556.0,205.0){\rule[-0.200pt]{0.400pt}{2.409pt}}
\put(556.0,1524.0){\rule[-0.200pt]{0.400pt}{2.409pt}}
\put(583.0,205.0){\rule[-0.200pt]{0.400pt}{2.409pt}}
\put(583.0,1524.0){\rule[-0.200pt]{0.400pt}{2.409pt}}
\put(605.0,205.0){\rule[-0.200pt]{0.400pt}{2.409pt}}
\put(605.0,1524.0){\rule[-0.200pt]{0.400pt}{2.409pt}}
\put(624.0,205.0){\rule[-0.200pt]{0.400pt}{2.409pt}}
\put(624.0,1524.0){\rule[-0.200pt]{0.400pt}{2.409pt}}
\put(641.0,205.0){\rule[-0.200pt]{0.400pt}{2.409pt}}
\put(641.0,1524.0){\rule[-0.200pt]{0.400pt}{2.409pt}}
\put(656.0,205.0){\rule[-0.200pt]{0.400pt}{4.818pt}}
\put(656.0,389.0){\rule[-0.200pt]{0.400pt}{275.830pt}}
\put(656.0,205.0){\rule[-0.200pt]{0.400pt}{4.818pt}}
\put(656,164){\makebox(0,0){ 100000}}
\put(656.0,1514.0){\rule[-0.200pt]{0.400pt}{4.818pt}}
\put(756.0,205.0){\rule[-0.200pt]{0.400pt}{2.409pt}}
\put(756.0,1524.0){\rule[-0.200pt]{0.400pt}{2.409pt}}
\put(815.0,205.0){\rule[-0.200pt]{0.400pt}{2.409pt}}
\put(815.0,1524.0){\rule[-0.200pt]{0.400pt}{2.409pt}}
\put(150.0,205.0){\rule[-0.200pt]{0.400pt}{320.156pt}}
\put(150.0,205.0){\rule[-0.200pt]{160.198pt}{0.400pt}}
\put(815.0,205.0){\rule[-0.200pt]{0.400pt}{320.156pt}}
\put(150.0,1534.0){\rule[-0.200pt]{160.198pt}{0.400pt}}
\put(49,869){\makebox(0,0){\rotatebox{90}{Time (sec)}}}
\put(482,103){\makebox(0,0){$N$}}
\sbox{\plotpoint}{\rule[-0.400pt]{0.800pt}{0.800pt}}%
\sbox{\plotpoint}{\rule[-0.200pt]{0.400pt}{0.400pt}}%
\put(655,369){\makebox(0,0)[r]{original}}
\sbox{\plotpoint}{\rule[-0.400pt]{0.800pt}{0.800pt}}%
\put(675.0,369.0){\rule[-0.400pt]{24.090pt}{0.800pt}}
\put(195,937){\usebox{\plotpoint}}
\multiput(196.41,937.00)(0.501,1.078){193}{\rule{0.121pt}{1.920pt}}
\multiput(193.34,937.00)(100.000,211.015){2}{\rule{0.800pt}{0.960pt}}
\multiput(296.41,1152.00)(0.501,0.917){193}{\rule{0.121pt}{1.664pt}}
\multiput(293.34,1152.00)(100.000,179.546){2}{\rule{0.800pt}{0.832pt}}
\multiput(396.41,1335.00)(0.502,0.904){103}{\rule{0.121pt}{1.640pt}}
\multiput(393.34,1335.00)(55.000,95.596){2}{\rule{0.800pt}{0.820pt}}
\put(195,937){\makebox(0,0){$\Diamond$}}
\put(295,1152){\makebox(0,0){$\Diamond$}}
\put(395,1335){\makebox(0,0){$\Diamond$}}
\put(450,1434){\makebox(0,0){$\Diamond$}}
\put(725,369){\makebox(0,0){$\Diamond$}}
\sbox{\plotpoint}{\rule[-0.200pt]{0.400pt}{0.400pt}}%
\put(655,328){\makebox(0,0)[r]{$T_{\rm wall}$}}
\sbox{\plotpoint}{\rule[-0.400pt]{0.800pt}{0.800pt}}%
\put(675.0,328.0){\rule[-0.400pt]{24.090pt}{0.800pt}}
\put(295,637){\usebox{\plotpoint}}
\multiput(296.41,637.00)(0.501,0.585){193}{\rule{0.121pt}{1.136pt}}
\multiput(293.34,637.00)(100.000,114.642){2}{\rule{0.800pt}{0.568pt}}
\multiput(396.41,754.00)(0.501,0.741){193}{\rule{0.121pt}{1.384pt}}
\multiput(393.34,754.00)(100.000,145.127){2}{\rule{0.800pt}{0.692pt}}
\multiput(496.41,902.00)(0.501,0.811){193}{\rule{0.121pt}{1.496pt}}
\multiput(493.34,902.00)(100.000,158.895){2}{\rule{0.800pt}{0.748pt}}
\multiput(596.41,1064.00)(0.501,0.847){193}{\rule{0.121pt}{1.552pt}}
\multiput(593.34,1064.00)(100.000,165.779){2}{\rule{0.800pt}{0.776pt}}
\multiput(696.41,1233.00)(0.501,0.933){195}{\rule{0.121pt}{1.689pt}}
\multiput(693.34,1233.00)(101.000,184.494){2}{\rule{0.800pt}{0.845pt}}
\put(295,637){\makebox(0,0){$\star$}}
\put(395,754){\makebox(0,0){$\star$}}
\put(495,902){\makebox(0,0){$\star$}}
\put(595,1064){\makebox(0,0){$\star$}}
\put(695,1233){\makebox(0,0){$\star$}}
\put(796,1421){\makebox(0,0){$\star$}}
\put(725,328){\makebox(0,0){$\star$}}
\sbox{\plotpoint}{\rule[-0.500pt]{1.000pt}{1.000pt}}%
\put(295,537){\usebox{\plotpoint}}
\multiput(295,537)(11.050,17.570){10}{\usebox{\plotpoint}}
\multiput(395,696)(11.100,17.538){9}{\usebox{\plotpoint}}
\multiput(495,854)(10.432,17.943){9}{\usebox{\plotpoint}}
\multiput(595,1026)(9.995,18.191){10}{\usebox{\plotpoint}}
\multiput(695,1208)(9.585,18.410){11}{\usebox{\plotpoint}}
\put(796,1402){\usebox{\plotpoint}}
\put(295,537){\makebox(0,0){$\star$}}
\put(395,696){\makebox(0,0){$\star$}}
\put(495,854){\makebox(0,0){$\star$}}
\put(595,1026){\makebox(0,0){$\star$}}
\put(695,1208){\makebox(0,0){$\star$}}
\put(796,1402){\makebox(0,0){$\star$}}
\sbox{\plotpoint}{\rule[-0.400pt]{0.800pt}{0.800pt}}%
\sbox{\plotpoint}{\rule[-0.200pt]{0.400pt}{0.400pt}}%
\put(655,287){\makebox(0,0)[r]{$T_{\rm reg}$}}
\sbox{\plotpoint}{\rule[-0.400pt]{0.800pt}{0.800pt}}%
\put(675.0,287.0){\rule[-0.400pt]{24.090pt}{0.800pt}}
\put(295,373){\usebox{\plotpoint}}
\multiput(296.41,373.00)(0.501,0.751){193}{\rule{0.121pt}{1.400pt}}
\multiput(293.34,373.00)(100.000,147.094){2}{\rule{0.800pt}{0.700pt}}
\multiput(396.41,523.00)(0.501,0.872){193}{\rule{0.121pt}{1.592pt}}
\multiput(393.34,523.00)(100.000,170.696){2}{\rule{0.800pt}{0.796pt}}
\multiput(496.41,697.00)(0.501,0.998){193}{\rule{0.121pt}{1.792pt}}
\multiput(493.34,697.00)(100.000,195.281){2}{\rule{0.800pt}{0.896pt}}
\multiput(596.41,896.00)(0.501,0.957){193}{\rule{0.121pt}{1.728pt}}
\multiput(593.34,896.00)(100.000,187.413){2}{\rule{0.800pt}{0.864pt}}
\multiput(696.41,1087.00)(0.501,1.003){195}{\rule{0.121pt}{1.800pt}}
\multiput(693.34,1087.00)(101.000,198.264){2}{\rule{0.800pt}{0.900pt}}
\put(295,373){\makebox(0,0){\tiny$\blacksquare$}}
\put(395,523){\makebox(0,0){\tiny$\blacksquare$}}
\put(495,697){\makebox(0,0){\tiny$\blacksquare$}}
\put(595,896){\makebox(0,0){\tiny$\blacksquare$}}
\put(695,1087){\makebox(0,0){\tiny$\blacksquare$}}
\put(796,1289){\makebox(0,0){\tiny$\blacksquare$}}
\put(725,287){\makebox(0,0){\tiny$\blacksquare$}}
\sbox{\plotpoint}{\rule[-0.500pt]{1.000pt}{1.000pt}}%
\put(295,366){\usebox{\plotpoint}}
\multiput(295,366)(12.354,16.678){9}{\usebox{\plotpoint}}
\multiput(395,501)(10.210,18.071){9}{\usebox{\plotpoint}}
\multiput(495,678)(9.870,18.259){11}{\usebox{\plotpoint}}
\multiput(595,863)(8.821,18.788){11}{\usebox{\plotpoint}}
\multiput(695,1076)(9.319,18.546){11}{\usebox{\plotpoint}}
\put(796,1277){\usebox{\plotpoint}}
\put(295,366){\raisebox{-.8pt}{\makebox(0,0){\tiny$\Box$}}}
\put(395,501){\raisebox{-.8pt}{\makebox(0,0){\tiny$\Box$}}}
\put(495,678){\raisebox{-.8pt}{\makebox(0,0){\tiny$\Box$}}}
\put(595,863){\raisebox{-.8pt}{\makebox(0,0){\tiny$\Box$}}}
\put(695,1076){\raisebox{-.8pt}{\makebox(0,0){\tiny$\Box$}}}
\put(796,1277){\raisebox{-.8pt}{\makebox(0,0){\tiny$\Box$}}}
\sbox{\plotpoint}{\rule[-0.400pt]{0.800pt}{0.800pt}}%
\sbox{\plotpoint}{\rule[-0.200pt]{0.400pt}{0.400pt}}%
\put(655,246){\makebox(0,0)[r]{$T_{\rm irr}$}}
\sbox{\plotpoint}{\rule[-0.400pt]{0.800pt}{0.800pt}}%
\put(675.0,246.0){\rule[-0.400pt]{24.090pt}{0.800pt}}
\put(295,403){\usebox{\plotpoint}}
\multiput(296.41,403.00)(0.501,0.821){193}{\rule{0.121pt}{1.512pt}}
\multiput(293.34,403.00)(100.000,160.862){2}{\rule{0.800pt}{0.756pt}}
\multiput(396.41,567.00)(0.501,0.816){193}{\rule{0.121pt}{1.504pt}}
\multiput(393.34,567.00)(100.000,159.878){2}{\rule{0.800pt}{0.752pt}}
\multiput(496.41,730.00)(0.501,0.857){193}{\rule{0.121pt}{1.568pt}}
\multiput(493.34,730.00)(100.000,167.746){2}{\rule{0.800pt}{0.784pt}}
\multiput(596.41,901.00)(0.501,0.902){193}{\rule{0.121pt}{1.640pt}}
\multiput(593.34,901.00)(100.000,176.596){2}{\rule{0.800pt}{0.820pt}}
\multiput(696.41,1081.00)(0.501,1.008){195}{\rule{0.121pt}{1.808pt}}
\multiput(693.34,1081.00)(101.000,199.248){2}{\rule{0.800pt}{0.904pt}}
\put(295,403){\makebox(0,0){$\blacktriangle$}}
\put(395,567){\makebox(0,0){$\blacktriangle$}}
\put(495,730){\makebox(0,0){$\blacktriangle$}}
\put(595,901){\makebox(0,0){$\blacktriangle$}}
\put(695,1081){\makebox(0,0){$\blacktriangle$}}
\put(796,1284){\makebox(0,0){$\blacktriangle$}}
\put(725,246){\makebox(0,0){$\blacktriangle$}}
\sbox{\plotpoint}{\rule[-0.500pt]{1.000pt}{1.000pt}}%
\put(295,286){\usebox{\plotpoint}}
\multiput(295,286)(9.549,18.429){11}{\usebox{\plotpoint}}
\multiput(395,479)(10.080,18.144){10}{\usebox{\plotpoint}}
\multiput(495,659)(11.895,17.009){8}{\usebox{\plotpoint}}
\multiput(595,802)(9.707,18.346){11}{\usebox{\plotpoint}}
\multiput(695,991)(8.315,19.017){12}{\usebox{\plotpoint}}
\put(796,1222){\usebox{\plotpoint}}
\put(295,286){\makebox(0,0){$\vartriangle$}}
\put(395,479){\makebox(0,0){$\vartriangle$}}
\put(495,659){\makebox(0,0){$\vartriangle$}}
\put(595,802){\makebox(0,0){$\vartriangle$}}
\put(695,991){\makebox(0,0){$\vartriangle$}}
\put(796,1222){\makebox(0,0){$\vartriangle$}}
\sbox{\plotpoint}{\rule[-0.200pt]{0.400pt}{0.400pt}}%
\put(150.0,205.0){\rule[-0.200pt]{0.400pt}{320.156pt}}
\put(150.0,205.0){\rule[-0.200pt]{160.198pt}{0.400pt}}
\put(815.0,205.0){\rule[-0.200pt]{0.400pt}{320.156pt}}
\put(150.0,1534.0){\rule[-0.200pt]{160.198pt}{0.400pt}}
\end{picture}
\caption{
	Wall-clock times for the original \nbody6 ($\Diamond$) and the \nbody{6\mathchar`-GPU} ($\star$) code
	for the interval $2\le t \le 4$.
	Partial times for the regular and irregular force are given
	by squares and triangles, respectively. Filled symbols and solid lines
	are for system A and open symbols and dotted lines for system B.
}
\end{figure}

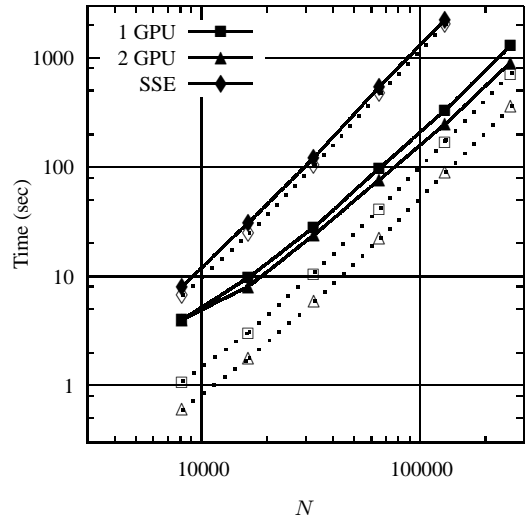
\begin{figure}
\setlength{\unitlength}{0.240900pt}
\ifx\plotpoint\undefined\newsavebox{\plotpoint}\fi
\sbox{\plotpoint}{\rule[-0.200pt]{0.400pt}{0.400pt}}%
\begin{picture}(826,810)(0,0)
\sbox{\plotpoint}{\rule[-0.200pt]{0.400pt}{0.400pt}}%
\put(140.0,169.0){\rule[-0.200pt]{2.409pt}{0.400pt}}
\put(815.0,169.0){\rule[-0.200pt]{2.409pt}{0.400pt}}
\put(140.0,204.0){\rule[-0.200pt]{2.409pt}{0.400pt}}
\put(815.0,204.0){\rule[-0.200pt]{2.409pt}{0.400pt}}
\put(140.0,221.0){\rule[-0.200pt]{165.016pt}{0.400pt}}
\put(140.0,221.0){\rule[-0.200pt]{4.818pt}{0.400pt}}
\put(120,221){\makebox(0,0)[r]{ 1}}
\put(805.0,221.0){\rule[-0.200pt]{4.818pt}{0.400pt}}
\put(140.0,272.0){\rule[-0.200pt]{2.409pt}{0.400pt}}
\put(815.0,272.0){\rule[-0.200pt]{2.409pt}{0.400pt}}
\put(140.0,341.0){\rule[-0.200pt]{2.409pt}{0.400pt}}
\put(815.0,341.0){\rule[-0.200pt]{2.409pt}{0.400pt}}
\put(140.0,376.0){\rule[-0.200pt]{2.409pt}{0.400pt}}
\put(815.0,376.0){\rule[-0.200pt]{2.409pt}{0.400pt}}
\put(140.0,392.0){\rule[-0.200pt]{165.016pt}{0.400pt}}
\put(140.0,392.0){\rule[-0.200pt]{4.818pt}{0.400pt}}
\put(120,392){\makebox(0,0)[r]{ 10}}
\put(805.0,392.0){\rule[-0.200pt]{4.818pt}{0.400pt}}
\put(140.0,444.0){\rule[-0.200pt]{2.409pt}{0.400pt}}
\put(815.0,444.0){\rule[-0.200pt]{2.409pt}{0.400pt}}
\put(140.0,512.0){\rule[-0.200pt]{2.409pt}{0.400pt}}
\put(815.0,512.0){\rule[-0.200pt]{2.409pt}{0.400pt}}
\put(140.0,547.0){\rule[-0.200pt]{2.409pt}{0.400pt}}
\put(815.0,547.0){\rule[-0.200pt]{2.409pt}{0.400pt}}
\put(140.0,564.0){\rule[-0.200pt]{165.016pt}{0.400pt}}
\put(140.0,564.0){\rule[-0.200pt]{4.818pt}{0.400pt}}
\put(120,564){\makebox(0,0)[r]{ 100}}
\put(805.0,564.0){\rule[-0.200pt]{4.818pt}{0.400pt}}
\put(140.0,615.0){\rule[-0.200pt]{2.409pt}{0.400pt}}
\put(815.0,615.0){\rule[-0.200pt]{2.409pt}{0.400pt}}
\put(140.0,684.0){\rule[-0.200pt]{2.409pt}{0.400pt}}
\put(815.0,684.0){\rule[-0.200pt]{2.409pt}{0.400pt}}
\put(140.0,719.0){\rule[-0.200pt]{2.409pt}{0.400pt}}
\put(815.0,719.0){\rule[-0.200pt]{2.409pt}{0.400pt}}
\put(140.0,735.0){\rule[-0.200pt]{4.818pt}{0.400pt}}
\put(420.0,735.0){\rule[-0.200pt]{97.564pt}{0.400pt}}
\put(140.0,735.0){\rule[-0.200pt]{4.818pt}{0.400pt}}
\put(120,735){\makebox(0,0)[r]{ 1000}}
\put(805.0,735.0){\rule[-0.200pt]{4.818pt}{0.400pt}}
\put(140.0,787.0){\rule[-0.200pt]{2.409pt}{0.400pt}}
\put(815.0,787.0){\rule[-0.200pt]{2.409pt}{0.400pt}}
\put(140.0,131.0){\rule[-0.200pt]{0.400pt}{2.409pt}}
\put(140.0,807.0){\rule[-0.200pt]{0.400pt}{2.409pt}}
\put(183.0,131.0){\rule[-0.200pt]{0.400pt}{2.409pt}}
\put(183.0,807.0){\rule[-0.200pt]{0.400pt}{2.409pt}}
\put(216.0,131.0){\rule[-0.200pt]{0.400pt}{2.409pt}}
\put(216.0,807.0){\rule[-0.200pt]{0.400pt}{2.409pt}}
\put(243.0,131.0){\rule[-0.200pt]{0.400pt}{2.409pt}}
\put(243.0,807.0){\rule[-0.200pt]{0.400pt}{2.409pt}}
\put(266.0,131.0){\rule[-0.200pt]{0.400pt}{2.409pt}}
\put(266.0,807.0){\rule[-0.200pt]{0.400pt}{2.409pt}}
\put(286.0,131.0){\rule[-0.200pt]{0.400pt}{2.409pt}}
\put(286.0,807.0){\rule[-0.200pt]{0.400pt}{2.409pt}}
\put(303.0,131.0){\rule[-0.200pt]{0.400pt}{2.409pt}}
\put(303.0,807.0){\rule[-0.200pt]{0.400pt}{2.409pt}}
\put(319.0,131.0){\rule[-0.200pt]{0.400pt}{130.809pt}}
\put(319.0,797.0){\rule[-0.200pt]{0.400pt}{4.818pt}}
\put(319.0,131.0){\rule[-0.200pt]{0.400pt}{4.818pt}}
\put(319,90){\makebox(0,0){ 10000}}
\put(319.0,797.0){\rule[-0.200pt]{0.400pt}{4.818pt}}
\put(422.0,131.0){\rule[-0.200pt]{0.400pt}{2.409pt}}
\put(422.0,807.0){\rule[-0.200pt]{0.400pt}{2.409pt}}
\put(483.0,131.0){\rule[-0.200pt]{0.400pt}{2.409pt}}
\put(483.0,807.0){\rule[-0.200pt]{0.400pt}{2.409pt}}
\put(525.0,131.0){\rule[-0.200pt]{0.400pt}{2.409pt}}
\put(525.0,807.0){\rule[-0.200pt]{0.400pt}{2.409pt}}
\put(558.0,131.0){\rule[-0.200pt]{0.400pt}{2.409pt}}
\put(558.0,807.0){\rule[-0.200pt]{0.400pt}{2.409pt}}
\put(586.0,131.0){\rule[-0.200pt]{0.400pt}{2.409pt}}
\put(586.0,807.0){\rule[-0.200pt]{0.400pt}{2.409pt}}
\put(609.0,131.0){\rule[-0.200pt]{0.400pt}{2.409pt}}
\put(609.0,807.0){\rule[-0.200pt]{0.400pt}{2.409pt}}
\put(628.0,131.0){\rule[-0.200pt]{0.400pt}{2.409pt}}
\put(628.0,807.0){\rule[-0.200pt]{0.400pt}{2.409pt}}
\put(646.0,131.0){\rule[-0.200pt]{0.400pt}{2.409pt}}
\put(646.0,807.0){\rule[-0.200pt]{0.400pt}{2.409pt}}
\put(662.0,131.0){\rule[-0.200pt]{0.400pt}{165.257pt}}
\put(662.0,131.0){\rule[-0.200pt]{0.400pt}{4.818pt}}
\put(662,90){\makebox(0,0){ 100000}}
\put(662.0,797.0){\rule[-0.200pt]{0.400pt}{4.818pt}}
\put(765.0,131.0){\rule[-0.200pt]{0.400pt}{2.409pt}}
\put(765.0,807.0){\rule[-0.200pt]{0.400pt}{2.409pt}}
\put(825.0,131.0){\rule[-0.200pt]{0.400pt}{2.409pt}}
\put(825.0,807.0){\rule[-0.200pt]{0.400pt}{2.409pt}}
\put(140.0,131.0){\rule[-0.200pt]{0.400pt}{165.257pt}}
\put(140.0,131.0){\rule[-0.200pt]{165.016pt}{0.400pt}}
\put(825.0,131.0){\rule[-0.200pt]{0.400pt}{165.257pt}}
\put(140.0,817.0){\rule[-0.200pt]{165.016pt}{0.400pt}}
\put(39,474){\makebox(0,0){\rotatebox{90}{Time (sec)}}}
\put(482,29){\makebox(0,0){$N$}}
\sbox{\plotpoint}{\rule[-0.400pt]{0.800pt}{0.800pt}}%
\sbox{\plotpoint}{\rule[-0.200pt]{0.400pt}{0.400pt}}%
\put(280,777){\makebox(0,0)[r]{ 1 GPU}}
\sbox{\plotpoint}{\rule[-0.400pt]{0.800pt}{0.800pt}}%
\put(300.0,777.0){\rule[-0.400pt]{24.090pt}{0.800pt}}
\put(289,324){\usebox{\plotpoint}}
\multiput(289.00,325.41)(0.790,0.501){125}{\rule{1.461pt}{0.121pt}}
\multiput(289.00,322.34)(100.968,66.000){2}{\rule{0.730pt}{0.800pt}}
\multiput(393.00,391.41)(0.652,0.501){151}{\rule{1.243pt}{0.121pt}}
\multiput(393.00,388.34)(100.420,79.000){2}{\rule{0.622pt}{0.800pt}}
\multiput(496.00,470.41)(0.553,0.501){179}{\rule{1.086pt}{0.121pt}}
\multiput(496.00,467.34)(100.746,93.000){2}{\rule{0.543pt}{0.800pt}}
\multiput(599.00,563.41)(0.566,0.501){175}{\rule{1.105pt}{0.121pt}}
\multiput(599.00,560.34)(100.705,91.000){2}{\rule{0.553pt}{0.800pt}}
\multiput(702.00,654.41)(0.509,0.501){195}{\rule{1.016pt}{0.121pt}}
\multiput(702.00,651.34)(100.892,101.000){2}{\rule{0.508pt}{0.800pt}}
\put(289,324){\makebox(0,0){\tiny$\blacksquare$}}
\put(393,390){\makebox(0,0){\tiny$\blacksquare$}}
\put(496,469){\makebox(0,0){\tiny$\blacksquare$}}
\put(599,562){\makebox(0,0){\tiny$\blacksquare$}}
\put(702,653){\makebox(0,0){\tiny$\blacksquare$}}
\put(805,754){\makebox(0,0){\tiny$\blacksquare$}}
\put(350,777){\makebox(0,0){\tiny$\blacksquare$}}
\sbox{\plotpoint}{\rule[-0.500pt]{1.000pt}{1.000pt}}%
\put(289,228){\usebox{\plotpoint}}
\multiput(289,228)(16.681,12.350){7}{\usebox{\plotpoint}}
\multiput(393,305)(15.331,13.991){6}{\usebox{\plotpoint}}
\multiput(496,399)(14.820,14.532){7}{\usebox{\plotpoint}}
\multiput(599,500)(14.464,14.885){8}{\usebox{\plotpoint}}
\multiput(702,606)(14.464,14.885){7}{\usebox{\plotpoint}}
\put(805,712){\usebox{\plotpoint}}
\put(289,228){\raisebox{-.8pt}{\makebox(0,0){\tiny$\Box$}}}
\put(393,305){\raisebox{-.8pt}{\makebox(0,0){\tiny$\Box$}}}
\put(496,399){\raisebox{-.8pt}{\makebox(0,0){\tiny$\Box$}}}
\put(599,500){\raisebox{-.8pt}{\makebox(0,0){\tiny$\Box$}}}
\put(702,606){\raisebox{-.8pt}{\makebox(0,0){\tiny$\Box$}}}
\put(805,712){\raisebox{-.8pt}{\makebox(0,0){\tiny$\Box$}}}
\sbox{\plotpoint}{\rule[-0.400pt]{0.800pt}{0.800pt}}%
\sbox{\plotpoint}{\rule[-0.200pt]{0.400pt}{0.400pt}}%
\put(280,736){\makebox(0,0)[r]{ 2 GPU}}
\sbox{\plotpoint}{\rule[-0.400pt]{0.800pt}{0.800pt}}%
\put(300.0,736.0){\rule[-0.400pt]{24.090pt}{0.800pt}}
\put(289,324){\usebox{\plotpoint}}
\multiput(289.00,325.41)(1.006,0.502){97}{\rule{1.800pt}{0.121pt}}
\multiput(289.00,322.34)(100.264,52.000){2}{\rule{0.900pt}{0.800pt}}
\multiput(393.00,377.41)(0.636,0.501){155}{\rule{1.217pt}{0.121pt}}
\multiput(393.00,374.34)(100.473,81.000){2}{\rule{0.609pt}{0.800pt}}
\multiput(496.00,458.41)(0.592,0.501){167}{\rule{1.147pt}{0.121pt}}
\multiput(496.00,455.34)(100.619,87.000){2}{\rule{0.574pt}{0.800pt}}
\multiput(599.00,545.41)(0.592,0.501){167}{\rule{1.147pt}{0.121pt}}
\multiput(599.00,542.34)(100.619,87.000){2}{\rule{0.574pt}{0.800pt}}
\multiput(702.00,632.41)(0.530,0.501){187}{\rule{1.049pt}{0.121pt}}
\multiput(702.00,629.34)(100.822,97.000){2}{\rule{0.525pt}{0.800pt}}
\put(289,324){\makebox(0,0){$\blacktriangle$}}
\put(393,376){\makebox(0,0){$\blacktriangle$}}
\put(496,457){\makebox(0,0){$\blacktriangle$}}
\put(599,544){\makebox(0,0){$\blacktriangle$}}
\put(702,631){\makebox(0,0){$\blacktriangle$}}
\put(805,728){\makebox(0,0){$\blacktriangle$}}
\put(350,736){\makebox(0,0){$\blacktriangle$}}
\sbox{\plotpoint}{\rule[-0.500pt]{1.000pt}{1.000pt}}%
\put(289,183){\usebox{\plotpoint}}
\multiput(289,183)(16.375,12.754){7}{\usebox{\plotpoint}}
\multiput(393,264)(15.705,13.570){6}{\usebox{\plotpoint}}
\multiput(496,353)(14.964,14.383){7}{\usebox{\plotpoint}}
\multiput(599,452)(14.605,14.747){7}{\usebox{\plotpoint}}
\multiput(702,556)(14.605,14.747){7}{\usebox{\plotpoint}}
\put(805,660){\usebox{\plotpoint}}
\put(289,183){\makebox(0,0){$\vartriangle$}}
\put(393,264){\makebox(0,0){$\vartriangle$}}
\put(496,353){\makebox(0,0){$\vartriangle$}}
\put(599,452){\makebox(0,0){$\vartriangle$}}
\put(702,556){\makebox(0,0){$\vartriangle$}}
\put(805,660){\makebox(0,0){$\vartriangle$}}
\sbox{\plotpoint}{\rule[-0.400pt]{0.800pt}{0.800pt}}%
\sbox{\plotpoint}{\rule[-0.200pt]{0.400pt}{0.400pt}}%
\put(280,695){\makebox(0,0)[r]{SSE}}
\sbox{\plotpoint}{\rule[-0.400pt]{0.800pt}{0.800pt}}%
\put(300.0,695.0){\rule[-0.400pt]{24.090pt}{0.800pt}}
\put(289,376){\usebox{\plotpoint}}
\multiput(289.00,377.41)(0.519,0.501){193}{\rule{1.032pt}{0.121pt}}
\multiput(289.00,374.34)(101.858,100.000){2}{\rule{0.516pt}{0.800pt}}
\multiput(393.00,477.41)(0.504,0.501){197}{\rule{1.008pt}{0.121pt}}
\multiput(393.00,474.34)(100.908,102.000){2}{\rule{0.504pt}{0.800pt}}
\multiput(497.41,578.00)(0.501,0.543){199}{\rule{0.121pt}{1.070pt}}
\multiput(494.34,578.00)(103.000,109.779){2}{\rule{0.800pt}{0.535pt}}
\multiput(600.41,690.00)(0.501,0.509){199}{\rule{0.121pt}{1.016pt}}
\multiput(597.34,690.00)(103.000,102.892){2}{\rule{0.800pt}{0.508pt}}
\put(289,376){\makebox(0,0){$\blacklozenge$}}
\put(393,476){\makebox(0,0){$\blacklozenge$}}
\put(496,578){\makebox(0,0){$\blacklozenge$}}
\put(599,690){\makebox(0,0){$\blacklozenge$}}
\put(702,795){\makebox(0,0){$\blacklozenge$}}
\put(350,695){\makebox(0,0){$\blacklozenge$}}
\sbox{\plotpoint}{\rule[-0.500pt]{1.000pt}{1.000pt}}%
\put(289,363){\usebox{\plotpoint}}
\multiput(289,363)(15.178,14.157){7}{\usebox{\plotpoint}}
\multiput(393,460)(14.394,14.953){8}{\usebox{\plotpoint}}
\multiput(496,567)(13.915,15.401){7}{\usebox{\plotpoint}}
\multiput(599,681)(14.325,15.020){7}{\usebox{\plotpoint}}
\put(702,789){\usebox{\plotpoint}}
\put(289,363){\makebox(0,0){$\lozenge$}}
\put(393,460){\makebox(0,0){$\lozenge$}}
\put(496,567){\makebox(0,0){$\lozenge$}}
\put(599,681){\makebox(0,0){$\lozenge$}}
\put(702,789){\makebox(0,0){$\lozenge$}}
\sbox{\plotpoint}{\rule[-0.200pt]{0.400pt}{0.400pt}}%
\put(140.0,131.0){\rule[-0.200pt]{0.400pt}{165.257pt}}
\put(140.0,131.0){\rule[-0.200pt]{165.016pt}{0.400pt}}
\put(825.0,131.0){\rule[-0.200pt]{0.400pt}{165.257pt}}
\put(140.0,817.0){\rule[-0.200pt]{165.016pt}{0.400pt}}
\end{picture}
\caption{
	Wall-clock times for the \nbody{6\text{--}SSE} code
	($\blacklozenge$, $\lozenge$) 
	and \nbody{6\text{--}GPU} code with one GPU
	({\tiny$\blacksquare$}, {\tiny$\Box$})
	and two GPUs
	($\blacktriangle$, $\vartriangle$)
	for the interval $2 \le t \le 4$.
	Solid lines with filled symbols 
	($\blacklozenge$, {\tiny$\blacksquare$}, $\blacktriangle$)
	are for the total times 
	and dotted lines with open symbols
	($\lozenge$, {\tiny$\Box$}, $\vartriangle$)
	for regular force times.
}
\end{figure}

For the basic performance tests we study isolated Plummer models in virial equilibrium
with equal-mass particles in the range $N = 8\text{--}256~{\rm k}$.
We use standard $N$-body units with total energy $E = -0.25$ and total mass $1$.
The benchmarks use two hardware systems, A and B, where the specifications
are summarized in \hyphenation{Table} Table~1.
Timing tests were first made for the older System~A.

In order to have a more balanced dynamical state, the computing times are measured
from $t = 2$ to $t = 4$.
The wall-clock time (in sec) as a function of $N$ is shown in Fig.~2 for both systems.
Also shown separately are the times for the regular and irregular force calculation.
As can be seen, these timings are quite similar for System A, especially in the
large-$N$ limit.
The minimum cost is achieved for particular choices of the upper limit of the neighbour
number, {\tt NBMAX}, which are close to values used in long-term simulations.
As for the remaining time expended by the FORTRAN part, it forms a diminishing
fraction of the total effort, with about 40 percent for $N=32~{\rm k}$ and 20
percent for $N=256~{\rm k}$.

Timing comparisons for the standard \nbody6 code have also been carried out.
Thus at the upper limit of $N = 24\,000$, the wall-clock time is 56 times the corresponding
time for System~A, with an asymptotic dependence $T_{\rm comp} \propto N^{1.8}$ 
in both cases.
Excluding the time consumption of the other parts, the performance of the regular force
calculation with System~A exceeds 1 TFLOPS.
As shown in Fig.~2, the CPU times for the irregular force with System~B are
almost a factor of 2 faster than for System~A when the slightly faster clock
is included.
This improvement is mainly due to the use of AVX instructions which became
available very recently.
More precise timings for System~B are presented in Table~2, as well as
time-step counts and average neighbour numbers for different $N$.
In comparison, the actual timings for $N=256~{\rm k}$ and System~A were
911, \,366 and 354 sec, respectively \footnote{Because of cache miss, there is a
degradation above $N = 64~{\rm k}$ for the irregular force calculation with AVX.}.
We also note that the average neighbour number scales
approximately as $N^{1/3}$.

The difference between using one and two GPUs is of interest.
In Fig.~3, we show timings on System A using one and two GPUs,
as well as a fully tuned CPU version with SSE and OpenMP
(\nbody{6\text{--}SSE}).
The wall-clock times are plotted in filled symbols with solid lines,
and times for the regular force in open symbols with dotted lines.
The performance gain in computing regular forces
from the SSE version to one GPU is about a factor of 10,
and this is doubled when going from one GPU to two GPUs.
However, the linear speed-up of the regular force calculations
does not always correspond to a good scalability of the total
simulation time because of the time consumed by the other parts.
For example with $N = 64~{\rm k}$, the regular force time
is reduced from 45 to 23 sec by using two GPUs, although
this becomes 99 and 77 sec for the total time.
Hence two separate simulations may be made simultaneously
on a dual GPU machine to keep the computing units busy,
provided $N$ is not too large.

The question of reproducibility in the multi-thread environment
is a difficult one, particularly in view of the
chaotic nature exhibited by the $N$-body problem.
On the other hand, the calculations are speeded up significantly by using
parallel OpenMP procedures.
At present, two important parallel treatments (lines 46--50 and 104--115 of
Listing~C1) are not thread-safe while some other FORTRAN parts are well behaved.
Consequently, strict reproducibility can be enforced by omitting these 
procedures at some loss of efficiency.
However, the timing tests were carried out with full optimization 
since
the early stage is essentially reproducible in any case.

The performance tests employed standard time-step parameters ($\eta = 0.02$)
for the irregular and regular force polynomials.
Typical relative energy errors for the time interval quoted are
$\Delta E/E \simeq 1.5 \times 10^{-7}$ ($N = 128$ and 256~k; also $N = 16$~k).
This compares favourably with values $4 \times 10^{-7}$ for the original
code ($N = 8$ and 16~k).
It should be noted that the intrinsic relative error in potential energy
evaluated on the GPU for efficiency is $1 \times 10^{-8}$ and hence on the
safe side.

Most $N$-body simulations, whether they be core collapse or substantial
evaporation, are concerned with long time-scales.
It is therefore desirable to assess the code behaviour for more advanced stages
of evolution with high core density which inevitably leads to binary formation and
strong interactions in compact subsystems.
At this stage the special treatments of close encounters in the form of KS and chain
regularization begin to play an important role.
The ejection of high-velocity members is a hallmark of an evolved dynamical state.
In general, the main energy errors are due to strong interactions, especially in
connection with switching to or from regularization procedures.
Even so, a study of core collapse for an equal-mass system with $N = 32~{\rm k}$
showed that the accumulated changes in total energy are surprisingly small,
amounting to $\sum \Delta E_j = -1\times 10^{-4}$ at minimum core radius.
A comparable drift in total energy was also seen in a test calculation well beyond
core collapse for $N = 16~{\rm k}$.
Hence the Hermite integration scheme has proved to possess excellent long-term stability.



\section{CONCLUSIONS}
We have presented new implementations for efficient integration of the $N$-body
problem with GPUs.
In the standard \nbody6 code, the regular force calculation dominates
the CPU time.
Consequently, the emphasis here has been on procedures for speeding up the
force calculation.
First the regular force evaluation was implemented on the GPU using the library
{\tt GPUNB} which also forms the neighbour list.
This procedure is ideally suited to massively parallel force calculations on GPUs
and resulted in significant gains.
However, a subsequent attempt to employ the GPU for the irregular force showed
that the overheads are too large.
It turned out that different strategies are needed for dealing with the regular
and irregular force components and this eventually led to the development of
the special library {\tt GPUIRR}.
The use of SSE and OpenMP speeded up this part such that the respective wall-clock
times are comparable for a range of particle numbers.

After the recent hardware with AVX support became available, 
the library was updated.
This led to additional speed-up of the irregular force calculation.
It is also essential that the regular force part scales well on multiple GPUs.
Thus in the future, further speed-up may be achieved by using four GPUs and
an octo-core CPU.
It should be noted that in the present scheme, the use of multiple GPUs only
benefits the regular force calculation which therefore scales well.
Although large $N$-body simulations are still quite expensive, we have
demonstrated that the regular part of the Ahmad--Cohen neighbour scheme is
well suited for use with GPU hardware.
Moreover, the current formulation of the \nbody{6\text{--}GPU} code performs
well for a variety of difficult conditions.

\section{ACKNOWLEDGMENTS}

Much of this work was done over the summers of 2007--2010 when K.~N. visited
the Institute of Astronomy, supported by the Visitor's Grant.
We also acknowledge helpful discussions with members of the \nbody6 community.
We thank Mark Gieles for making System B available.
The full source code is freely available on
{\tt http://www.ast.cam.ac.uk/research/nbody/}.

\label{lastpage}

\appendix
\section{Glossary}
\begin{description}
\item[\bf GCC] GNU Compiler Collection, including {\tt gcc}, {\tt g++}, {\tt gfortran} 
	and other languages.
\item[\bf API] Application Programing Interface.
	This provides definitions or prototypes of FORTRAN subroutines or C functions.
\item[\bf SIMD] Single Instruction Multiple Data.
	A model of parallel computer where an operation such as addition and
	multiplication is performed for multiple data.
\item[\bf SSE]  Streaming SIMD Extensions. 
	Additional instruction set for x86.
	Four words single precision floating point operations on 128-bit registers are supported.
\item[\bf AVX ]  Advanced Vector eXtensions. 
	Further enlargement of SSE.
	Eight words single precision or four words double precision 
	floating point operations on 256-bit registers.
\item[\bf CUDA]  Compute Unified Device Architecture.
	A frame-work for general purpose computing on NVIDA GPUs,
	including language, compiler, run-time library and device driver.
\item[\bf OpenMP] A standard to utilize multiple
	processors on shared memory from high-level languages,
	provided as directives of the language.
\item[\bf Thread] A unit of parallel execution.
	Threads of CPU execute different contexts, while
	a cluster of threads of GPU executes the same context for different data.
\item[\bf Thread-Block] A number of GPU threads which share the same context.
\item[\bf Kernel] A short program submitted and executed on GPU.
\item[\bf {\itshape i}-particle] A particle which feels the gravitational force, 
	named from the outer loop index.
\item[\bf {\itshape j}-particle] A particle which is a source of the gravitational force, 
	named from the inner loop index.
\item[\bf {\itshape i}-parallelism] Parallelism for the outer loop.
	Forces on different particles are calculated in parallel.
\item[\bf {\itshape j}-parallelism] Parallelism for the inner loop.
	Forces from different particles are calculated in parallel,
	and summed later.
\end{description}

\lstloadlanguages{C, Fortran}
\lstset{%
	language={C},%
	morekeywords={float3, __device__},
	basicstyle={\scriptsize\tt},%
	identifierstyle={},%
	keywordstyle={\scriptsize\bfseries},%
	ndkeywordstyle={},%
	stringstyle={\ttfamily},
	commentstyle={\scriptsize\rmfamily\itshape},
	frame={single},%
	breaklines=false,%
	columns=[l]{fixed},%
	numbers=left,%
	tabsize=4,%
	xleftmargin=10pt,%
	numberstyle={\scriptsize},%
	stepnumber=1,%
	lineskip=0.15ex,%
}

\renewcommand{\lstlistingname}{\small Listing}
\renewcommand\thelstlisting{\thesection\arabic{lstlisting}}

\section{CUDA codes}
\setcounter{lstlisting}{0}
In Listing B1 and B2, we show the innermost regular force kernel
written in CUDA C++ as well as some definitions of data structures.
Each CUDA thread holds one $i$-particle in its registers,
and evaluates a force from particle $j$ and accumulates it.
The functions qualified with {\tt \_\_device\_\_} are forced
to be in-line. The argument {\tt nblist} is a pointer to
the device memory.

\begin{lstlisting}[caption={Definitions of structures.}]
struct Jparticle{
	float3 pos;
	float  mass;
	float3 vel;
	float  pad; // 64-byte
};
struct Iparticle{
	float3 pos;
	float  h2;
	float3 vel;
	float  dtr; // 64-byte
};
struct Force{
	float3 acc;
	float  pot;
	float3 jrk;
	int    nnb; //  64-byte
};
\end{lstlisting}


\begin{lstlisting}[caption={The innermost gravity calulation.}]
// num of neib per block, must be power of 2
#define NNB_PER_BLOCK 256 

__device__ void dev_gravity(
		const int        jidx,
		const Iparticle &ip, 
		const Jparticle &jp, 
		Force           &fo,
		uint16           nblist[])
{
	float dx  = jp.pos.x - ip.pos.x;
	float dy  = jp.pos.y - ip.pos.y;
	float dz  = jp.pos.z - ip.pos.z;
	float dvx = jp.vel.x - ip.vel.x;
	float dvy = jp.vel.y - ip.vel.y;
	float dvz = jp.vel.z - ip.vel.z;
	float dxp = dx + ip.dtr * dvx;
	float dyp = dy + ip.dtr * dvy;
	float dzp = dz + ip.dtr * dvz;

	float r2  = dx *dx  + dy *dy  + dz *dz;
	float r2p = dxp*dxp + dyp*dyp + dzp*dzp;
	float rv  = dx *dvx + dy *dvy + dz *dvz;

	float rinv1 = rsqrtf(r2);
	if(fminf(r2, r2p)  < ip.h2){
		// address to avoid buffer overflow
		int addr = fo.nnb & (NNB_PER_BLOCK -1);
		nblist[addr] = (uint16)jidx;
		fo.nnb++;
		rinv1 = 0.f;
	}
	float rinv2 = rinv1 * rinv1;
	float mrinv1 = jp.mass * rinv1;
	float mrinv3 = mrinv1 * rinv2;
	float alpha  = -3.f * rv * rinv2;
	
#ifdef POTENTIAL
	fo.pot += mrinv1;
#endif
	fo.acc.x += mrinv3 * dx;
	fo.acc.y += mrinv3 * dy;
	fo.acc.z += mrinv3 * dz;
	fo.jrk.x += mrinv3 * (dvx + alpha * dx);
	fo.jrk.y += mrinv3 * (dvy + alpha * dy);
	fo.jrk.z += mrinv3 * (dvz + alpha * dz);
}
\end{lstlisting}
In lines 17--19, 22 and 26, we can see the additional operations
for the velocity criterion (5 mul, 6 add, 1 min) discussed in Section 3.2.

\section{FORTRAN codes}
\setcounter{lstlisting}{0}
It is instructive to examine the program flow for the treatment of the
regular and irregular force during one block-step.
We have copied the relevant FORTRAN parts, omitting some extra features, and
display a complete cycle in the general case.
Consider the situation when the next block-step time, denoted by {\tt TMIN},
has been determined at the end of the previous step (as shown below).
All particles due for consideration up to a certain time are contained in
{\tt LISTQ}, and the new block-step members {\tt NXTLEN} which satisfy equation (\ref{eq:Lact})
are saved in the list
array {\tt NXTLST} after the first call to {\tt INEXT}.
Procedures for advancing KS and chain solutions which usually follow here
are omitted for simplicity.
The essential symbols are defined in Table~C1, where some are fixed parameters.
For completeness, we also include the {\tt OPEN} and {\tt CLOSE} statements,
normally only used at the start and end.
With this preamble and the definitions of Table~C1 the code section can now
be inspected with assistance from the explanatory comments.
The purpose of most special procedures have already been discussed in the text and
the naming is intended to be descriptive.

\begin{table}
\caption{List of symbols.}
\begin{center}
\begin{small}
\begin{tabular}{>{\fontsize{7pt}{0pt}\tt}l >{\fontsize{7pt}{0pt}\tt}l >{\fontsize{7pt}{0pt}}l}
\hline
\multicolumn{1}{l}{Symbol    } &
\multicolumn{1}{l}{Data type } &
\multicolumn{1}{l}{Definition} \\
\hline
LMAX    &  PARAMETER       & Size of compiled neighbour arrays \\
NBMAX   &  PARAMETER       & Maximum neighbour number \\
NIMAX   &  PARAMETER       & Block-size of regular force loop \\
NMAX    &  PARAMETER       & Maximum particle number \\
NPACT   &  PARAMETER       & Limit for active particle prediction \\
NPMAX   &  PARAMETER       & Parallel irregular integration \\
IFIRST  &  INTEGER         & Array index of first single particle \\
NFR     &  INTEGER         & Number of regular force calculations \\
NQ      &  INTEGER         & Membership of {\tt LISTQ} \\
NTOT    &  INTEGER         & Number of single particles and KS pairs \\
LISTQ   &  INT (NMAX)      & List of particles due before a given time \\
NXTLEN  &  INT (NMAX)      & Length of current block-step list \\
NXTLST  &  INT (NMAX)      & List of block-step members \\
RS      &  REAL(NMAX)      & Radius of individual neighbour sphere \\
STEPR   &  REAL(NMAX)      & Regular time-step \\
T0      &  REAL(NMAX)      & Time of last irregular force calculation \\
TMIN    &  REAL(NMAX)      & New block-time \\
TNEW    &  REAL(NMAX)      & Next irregular force time \\
X       &  REAL(3,NMAX)    & Predicted coordinates \\
X0      &  REAL(3,NMAX)    & Corrected coordinates \\
XDOT    &  REAL(3,NMAX)    & Predicted velocities \\
X0DOT   &  REAL(3,NMAX)    & Corrected velocities \\
\hline
\end{tabular}
\end{small}
\end{center}
\end{table}

\lstset{
	language={Fortran},%
	basicstyle={\tiny\tt},%
	morecomment=[f][\tiny\it][0]{*},%
	commentstyle={\tiny\it},%
}
\begin{lstlisting}[caption={extract of {\ttfamily intgrt.omp.f}.}]
*       Declare typical parameters for N=64k.
      PARAMETER(NPACT=150, NPMAX=16, NIMAX=1024)
*       Open the regular and irregular libraries.
      CALL GPUNB_OPEN(NTOT)
      CALL GPUIRR_OPEN(NTOT,LMAX)
*
*       Find all particles in next block (TNEW = TMIN) and set TIME.
      CALL INEXT(NQ,LISTQ,TMIN,NXTLEN,NXTLST)
      TIME = TMIN
*
*       Form lists of candidates for new irregular and regular force.
      NFR = 0
      DO 28 L = 1,NXTLEN
          J = NXTLST(L)
          IF (TNEW(J).GE.T0R(J) + STEPR(J)) THEN
              NFR = NFR + 1
              IREG(NFR) = J
              IRR(L) = J
          ELSE
              IRR(L) = 0
          END IF
   28 CONTINUE
* 
*       Decide between predicting <= NPACT active (NFR=0) or all particles.
      IF (NXTLEN.LE.NPACT.AND.NFR.EQ.0) THEN
          CALL GPUIRR_PRED_ACT(NXTLEN,NXTLST,TIME)
      ELSE
          CALL GPUIRR_PRED_ALL(IFIRST,NTOT,TIME)
      END IF
*
*       Evaluate new irregular forces & derivatives in the GPUIRR library.
      CALL GPUIRR_FIRR_VEC(NXTLEN,NXTLST,GF,GFD)
*
*       Choose between standard and parallel irregular integration.
      IF (NXTLEN.LE.NPMAX) THEN
*
*       Correct the irregular steps sequentially.
      DO 48 II = 1,NXTLEN
          I = NXTLST(II)
          CALL NBINT(I,IRR(II),GF(1,II),GFD(1,II))
   48 CONTINUE
*
      ELSE
*
*       Perform irregular correction in parallel.
!$omp parallel do private(II, I)
      DO 50 II = 1,NXTLEN
          I = NXTLST(II)
          CALL NBINTP(I,IRR(II),GF(1,II),GFD(1,II))
   50 CONTINUE
!$omp end parallel do
      END IF
*
*       Check regular force updates (NFR members on block-step).
      IF (NFR.GT.0) THEN
*
*       Predict all particles (except TPRED=TIME) in C++ on host.
      CALL CXVPRED(IFIRST,NTOT,TIME,T0,X0,X0DOT,F,FDOT,X,XDOT,TPRED)
*       Send all single particles and c.m. bodies to the GPU.
      NN = NTOT - IFIRST + 1
      CALL GPUNB_SEND(NN,BODY(IFIRST),X(1,IFIRST),XDOT(1,IFIRST))
*
*       Perform regular force loop (blocks of NIMAX=1024).
      JNEXT = 0
      DO 55 II = 1,NFR,NIMAX
          NI = MIN(NFR-JNEXT,NIMAX)
*       Copy neighbour radius, STEPR and state vector for each block.
!$omp parallel do private(LL, I, K)
          DO 52 LL = 1,NI
              I = IREG(JNEXT+LL)
              H2I(LL) = RS(I)**2
              DTR(LL) = STEPR(I)
              DO 51 K = 1,3
                  XI(K,LL) = X(K,I)
                  VI(K,LL) = XDOT(K,I)
   51         CONTINUE
   52     CONTINUE
!$omp end parallel do
*
*       Evaluate forces, derivatives and neighbour lists for new block.
          CALL GPUNB_REGF(NI,H2I,DTR,XI,VI,GPUACC,GPUJRK,GPUPHI,LMAX,
     &                                                   NBMAX,LISTGP)
*       Copy neighbour lists from the GPU.
!$omp parallel do private(LL, I, ITEMP, NNB, L1, L)
          DO 56 LL = 1,NI
              I = IREG(JNEXT + LL)
              NNB = LISTGP(1,LL)
              L1 = 1
              DO 53 L = 2,NNB+1
*       Note GPU address starts from 0 (hence add IFIRST to neighbour list).
                  ITEMP = LISTGP(L,LL) + IFIRST
                  IF (ITEMP.NE.I) THEN
                      L1 = L1 + 1
                      LISTGP(L1,LL) = ITEMP
                  END IF
   53         CONTINUE
              LISTGP(1,LL) = L1 - 1
              CALL GPUIRR_SET_LIST(I,LISTGP(1,LL))
   56     CONTINUE
!$omp end parallel do
*
*       Evaluate current irregular forces by vector procedure.
          CALL GPUIRR_FIRR_VEC(NI,IREG(II),GF(1,1),GFD(1,1))
!$omp parallel do private(LL, I, LX)
          DO 57 LL = 1,NI
              I = IREG(JNEXT+LL)
*       Send new irregular force and perform Hermite corrector.
              CALL GPUCOR(I,XI(1,LL),VI(1,LL),GPUACC(1,LL),GPUJRK(1,LL),
     &                               GF(1,LL),GFD(1,LL),LISTGP(1,LL),LX)
*       Update neighbour lists in GPUIRR library (only if changed: LX > 0).
              IF (LX.GT.0) THEN
                  CALL GPUIRR_SET_LIST(I,LIST(1,I))
              END IF
   57     CONTINUE
!$omp end parallel do
          JNEXT = JNEXT + NI
   55 CONTINUE
      END IF
*
*       Determine next block time (note STEP may shrink in GPUCOR).
      TMIN = 1.0D+10
      DO 60 L = 1,NXTLEN
          I = NXTLST(L)
          IF (TNEW(I).LT.TMIN) THEN
              TMIN = TNEW(I)
          END IF
   60 CONTINUE
*
*       Copy current coordinates & velocities from corrected values.
!$omp parallel do private(I, L, K)
      DO 70 L = 1,NXTLEN
          I = NXTLST(L)
          DO 65 K = 1,3
              X(K,I) = X0(K,I)
              XDOT(K,I) = X0DOT(K,I)
   65     CONTINUE
*       Send corrected active particles to GPUIRR library.
          CALL GPUIRR_SET_JP(I,X0(1,I),X0DOT(1,I),F(1,I),FDOT(1,I),
     &                                            BODY(I),T0(I))
   70 CONTINUE
!$omp end parallel do
      CALL GPUNB_CLOSE
      CALL GPUIRR_CLOSE
\end{lstlisting}

\end{document}